\documentstyle[epsfig]{mn}

\title{Pulsation of the $\delta$~Scuti star $\theta^2$~Tau:
New multisite photometry and modelling of instability}
  
\author[M. Breger et al.]
{M.~Breger,$^1$ A.~A.~Pamyatnykh,$^{1,2,3}$ W.~Zima,$^1$
R.~Garrido,$^4$ G.~Handler,$^1 $ P.~Reegen$^1$\\
$^1$Astronomisches Institut der Universit\"at Wien, T\"urkenschanzstr. 17,
A--1180 Wien, Austria\\
$^2$Copernicus Astronomical Center, Bartycka 18, 00-716 Warsaw, Poland\\
$^3$Institute of Astronomy, Russian Academy of Sciences, Pyatnitskaya Str. 48,
109017 Moscow, Russia\\
$^4$Instituto de Astrofisica de Andalucia, CSIC, Apdo. 3004, E-18080 Granada, Spain\\
}
\date{Accepted 2002 month day.
      Received 2002 month day;
      in original form 2002 month date}

\begin{document}
\maketitle
 
\begin{abstract}

The results of a multisite photometric campaign of $\theta^2$ Tau are reported.
This binary system consists of evolved and main-sequence A stars inside the instability strip.
The 12th Delta Scuti Network campaign included 152 hours of high-precision photometry
obtained at four observatories. This leads to the derivation of 11 frequencies of
pulsation in the 10.8 to 14.6~cd$^{-1}$ range. These frequencies confirm the
results from previous Earth-based (1982--1986) as well as satellite (2000) photometry,
although amplitude variability on a time scale of several years is present.

We show that at least two high frequencies (26.18 and 26.73~cd$^{-1}$) are also
present in $\theta^2$~Tau. Arguments are given that these high frequencies originate
in the main-sequence companion and are not combination frequencies,
f$_i$+f$_j$, from the primary.

Models for both the primary and the secondary components
were checked for the instability against radial and nonradial
oscillations.  All hot models of the primary with
$T_{\rm{eff}} > 8000$\,K are stable in the observed frequency range.
The best fit between the theoretical and observed frequency ranges is
achieved for models with $T_{\rm{eff}}\approx 7800$\,K (or slightly higher),
in agreement with photometric calibrations. The
instability range spans two or three radial orders in the range $p_4$ to $p_6$
for radial modes. 
Post-main-sequence models (with or without overshooting) are preferable
for the primary, but main-sequence models with overshooting cannot be excluded.
For the less luminous secondary component the instability range is wider
and spans 5 to 7 radial orders from $p_2$ to $p_8$.  The observed
frequencies lie around radial modes $p_5$ to $p_6$. The main uncertainties
of these results are caused by a simple treatment of
the convective flux in the hydrogen ionization zone.

\end{abstract}


\section{Introduction}

A number of lengthy observing campaigns covering individual $\delta$~Scuti variables
have shown
that the majority of these pulsating variables on and near the main sequence
pulsate with a large number of simultaneously excited nonradial p modes.
Furthermore, long-term variability of the pulsation amplitudes of these
nonradial modes with a time scale of years has been discovered for many,
but not all of these nonradial pulsators. The Delta Scuti Network, hereafter
called DSN, specializes in multisite observations of these stars. The network
is a collaboration of astronomers located at observatories spaced around the globe in
order to avoid regular, daily observing gaps. So far, 21 campaigns have been carried
out. The most recent campaign covered BI~CMi, for which 29 frequencies of pulsation
derived from 1024 hours (177 nights) of photometry were discovered (Breger et al. 2002).

The variability of the star $\theta^2$~Tau was discovered by Horan (1977, 1979) and confirmed
by Duerbeck (1978). In order to study the multiple frequencies, the Delta Scuti Network undertook
two campaigns (Breger et al. 1987, 1989: the latter is referred to as Paper I).
The data could be supplemented by additional
measurements by Kovacs \& Paparo (1989). Five frequencies of pulsation were detected and
interpreted to be due to nonradial pulsation. The size and distribution of the frequencies
suggested p modes with values of $\ell$ = 0 to 2.
A mode with a higher $\ell$ value was also detected spectroscopically
by Kennelly \& Walker (1996). Since the original analysis of the
photometric data, new theoretical as well as experimentally determined (Breger et al. 1993)
statistical criteria for the acceptance or rejection of additional
modes extracted from multisite photometric data have become available.
A re-examination of the previous data obtained in the years 1982
to 1986 (Breger 2002, hereafter referred to as Paper II) led to the discovery
of ten frequencies in the 1986 DSN data.
Nine of these frequencies are also present in the list of
twelve frequencies determined during a remarkable
study of $\theta^2$ Tau from space (Poretti et al. 2002,
hereafter referred to as Paper III): during 2000 August,
$\theta^2$ Tau was monitored extensively with the star tracker camera on the Wide-Field
Infrared Explorer satellite (WIRE).

$\theta^2$~Tau is a 140.728d binary system with known orbital elements (Ebbighausen 1959,
Torres, Stefanik \& Latham 1997). The two components have similar temperatures. The primary
component is evolved (A7III), while the secondary is fainter by 1.10 mag
(Peterson, Stefanik \& Latham 1993) and still on the main sequence. Both stars are
situated inside the instability strip. It was shown in Paper I that the dominant
pulsation modes originate in the primary component because the
predicted orbital light-time effects for the primary match the observed
shifts in the light curves of up to several minutes. This confirms an expectation
from the values of the frequencies (10 -- 15~cd$^{-1}$),
which are incompatible with those of main-sequence $\delta$~Scuti stars.

\section{New ground-based measurements}

\begin{table*}
\caption{Journal of photoelectric measurements of $\theta^2$ Tau
obtained during the 12th Delta Scuti Network campaign}
\begin{flushleft}
\begin{tabular}{lll|lll|lll}
\hline
\noalign{\smallskip}
Start& Length & Observatory & Start & Length & Observatory & Start& Length & Observatory \\
 HJD & hours & & HJD & hours & &HJD & hours \\
\noalign{\smallskip}
\hline
\noalign{\smallskip}
244 9000+ &&&244 9000+&&&244 9000+\\
\noalign{\smallskip}
667.186	&	0.90	&	XL	&	682.347	&	8.69	&	SNO	&	687.761	&	5.05	&	McD	\\
670.697	&	7.77	&	Lo	&	683.337	&	9.25	&	SNO	&	688.243	&	0.85	&	XL	\\
671.678	&	8.18	&	Lo	&	684.263	&	1.15	&	XL	&	688.919	&	1.16	&	McD	\\
674.222	&	2.93	&	XL	&	684.335	&	9.10	&	SNO	&	688.986	&	8.39	&	XL	\\
678.776	&	0.79	&	McD	&	684.651	&	1.61	&	McD	&	689.816	&	2.58	&	McD	\\
679.637	&	5.90	&	McD	&	684.671	&	7.70	&	Lo	&	689.999	&	0.68	&	XL	\\
680.562	&	3.80	&	SNO	&	685.065	&	6.50	&	XL	&	690.426	&	6.41	&	SNO	\\
680.957	&	0.99	&	McD	&	685.649	&	8.24	&	Lo	&	691.018	&	7.38	&	XL	\\
681.101	&	5.65	&	XL	&	686.640	&	8.56	&	Lo	&	691.590	&	2.60	&	McD	\\
681.328	&	9.45	&	SNO	&	686.779	&	4.75	&	McD	&	692.299	&	4.45	&	SNO	\\
\noalign{\smallskip}      
\hline
\noalign{\smallskip}
\end{tabular}
\newline
Lo: Lowell Observatory, USA; McD: McDonald Observatory, USA; SNO: Sierra Nevada Observatory, Spain\\
XL: Xing-Long Observatory, China\\
\end{flushleft}
\end{table*}

In order to eliminate the serious aliasing caused by regular observing gaps,
a multisite campaign was organized utilizing the Delta Scuti Network (DSN
campaign 12). During 1994 November and December, $\theta^2$ Tau was measured
photometrically with the Three-Star-Technique (Breger 1993) at a number of observatories
spaced around the globe. Weather conditions, unfortunately, were unfavorable so that
data from a number of Asian Observatories were not accurate enough to be included
in the present analysis. 152 hours of photometry at four other observatories could be used.
These observatories were:

(i) Lowell: The photoelectric photometer attached to the 0.8m reflector at Lowell
Observatory, Flagstaff, USA, was used with the Str{\"o}mgren $v$ and $y$ filters
together with a neutral density filter. The observer was W. Zima.

(ii) McDonald Observatory: The 0.9m telescope was used with the $y$ filter. 
The observer was G. Handler.

(iii) Sierra Nevada Observatory: The 0.9m telescope was used together with
$v$ and $y$ filters. The observers were R. Garrido and F. Beichbuchner.

(iv) The 0.6m reflector at the XingLong station of Beijing Astronomical Observatory, China,
was used for 9 nights with the $V$ filter. The observers were Li Zhiping, Zhou Aiying,
and Yang Dawei. More details can be found in Li et al. (1997a),
where the Chinese data have already been published,
and Li et al. (1997b). We have adopted the Chinese data as published
except for the slight editing of a few (obviously) deviant points.

A journal of the photoelectric measurements used is given in Table 1.

\begin{figure}
\centering
\includegraphics[width=86mm,clip]{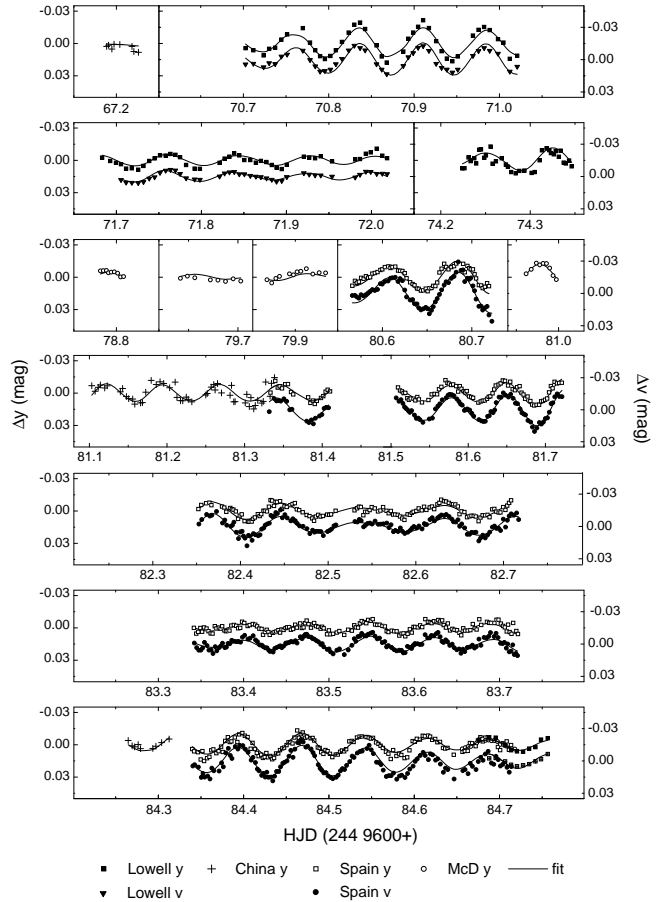}
\caption{Light curves showing the first half of the 1994 measurements of $\theta^2$ Tau. The
curves drawn represent the 13-frequency fit found in a later section.}
\end{figure}

\begin{figure}
\centering
\includegraphics[width=86mm,clip]{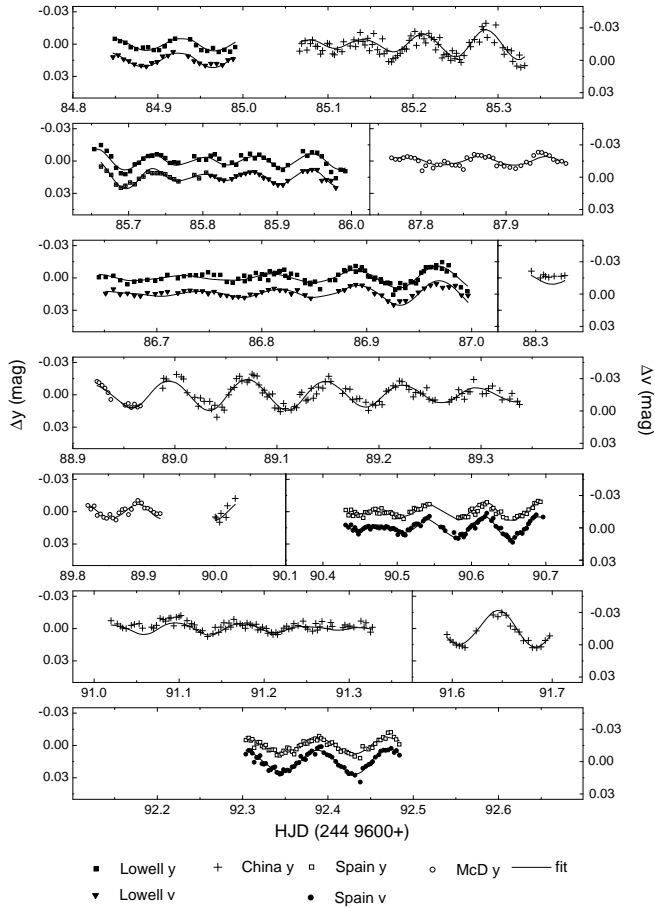}
\caption{Light curves showing the second half of the 1994 measurements of $\theta^2$ Tau.}
\end{figure}

All observatories used the same two comparison stars as during the previous DSN
campaigns of $\theta^2$ Tau, viz., HR~1422 and HR~1428. No evidence for any variability
of these comparison stars was found. We could combine the $V$ and $y$ data, because the $y$
magnitudes are defined to be equal to the $V$ magnitudes. After the standard photometric reductions
relative to the two comparison stars, the data from the different observatories
needed to be combined. For the initial multifrequency analyses,
this was done by subtracting the average relative magnitude of $\theta^2$ Tau at each observatory and
then combining the data. Since
the multifrequency analyses utilized the fitting of sine curves to the data,
the zero-points could then be further adjusted from the values of zero-point
of these fits. The times of observation were converted to Heliocentric Julian Date
(HJD). We also calculated the light time corrections for the orbital motion of the primary
stellar component (see Fig. 6 of Paper I). During this short campaign of 25d,
the light time correction is very small and practically constant. Therefore, we have not
applied such a correction to the present data.
The extinction coefficients were derived separately for each night by using  
the two comparison stars. For a few short data sets, average extinction
coefficients had to be applied. 

Figures 1 and 2 show the observed light variations of $\theta^2$ Tau together
with the 13-frequency fit derived in later sections. The new data
will be made available in the Communications in Asteroseismology.

\section{Methods used to determine multiple frequencies from the light curves}

The pulsation frequency analyses were performed with a package of computer
programs with single-frequency and multiple-frequency techniques
(PERIOD98, Sperl 1998). These programs utilize Fourier as well as
multiple-least-squares algorithms. The latter technique fits a number of
simultaneous sinusoidal variations to the observed light variability
and does not rely
on prewhitening. For the purposes of presentation and initial searches, however, prewhitening is
required if the low-amplitude modes are to be seen. Therefore, in the presentation
of the results (see below), the various
power spectra are presented as a series of panels, each with additional
frequencies removed relative to the panel above.

Two observatories also provided measurements in a second filter, viz., the Str{\"o}mgren $v$ filter.
In this filter, the amplitudes of $\theta^2$~Tau are higher by about 30\% relative those in the $y$ filter.
In power spectra, the location of the peaks in frequency are not affected
by mixing data with different amplitudes. Consequently, for the power spectra
we have used the data from all filters. The slight phase shift up
to 10$\degr$ in the light curves of $\delta$~Scuti stars
between the different
filters was ignored, since numerical simulations show that for $\delta$~Scuti stars
the power spectra are not affected by such small shifts.

However, for prewhitening, the data from the two filters had to be treated separately.
The multiperiodic fitting of sinusoids to the data relies on the minimization
of the residuals between the fit and the observations and requires correct amplitudes. This
problem was solved by computing separate solutions for the two colors, which could then
be prewhitened.

One of the most important questions in the examination of multiperiodicity
concerns the decision as to which of the detected peaks in the power spectrum can  
be regarded as variability intrinsic to the star. Due to the presence of  
nonrandom errors in photometric observations and because of observing gaps, the  
predictions of standard statistical false-alarm tests give answers which are considered  
by us to be overly optimistic. In a previous paper (Breger et al. 1993) we have
argued that a ratio of amplitude signal/noise = 4.0 provides a useful  
criterion for judging the reality of a peak. This corresponds to a power
signal/noise ratio of 12.6. Subsequent analyses comparing
independent data sets have confirmed that this criterion is an excellent predictor of
intrinsic vs. possible noise peaks, as long as it is not applied to very small data sets
or at low frequencies, where the errors of measurement are far from random.
In the present study, the noise
was calculated by averaging the amplitudes (oversampled by a factor of 20)  
over 5~cd$^{-1}$ regions centered around the frequency under consideration.

\section{The main pulsation frequencies of $\theta^2$ Tau during 1994}

\begin{figure}
\centering
\includegraphics[width=86mm,clip]{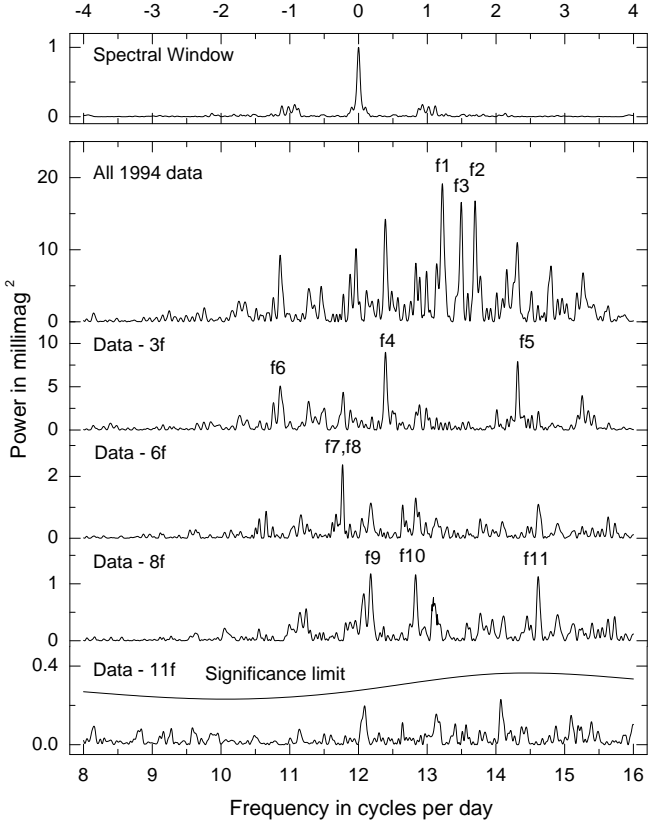}
\caption{Power spectra of the 1994 photometry of $\theta^2$~Tau.
The top panel shows the spectral window, while the other panels
present the power spectra before and after prewhitening a given
number of frequencies. See text for a discussion of significance levels.}
\end{figure}

The spectral window of the 1994 data (Fig.~3) is quite clean because the data were
collected on three continents, thereby avoiding serious regular day-time observing
gaps. Nevertheless, the 1~cd$^{-1}$ aliasing should still be kept in mind, since modes with
very small amplitudes might hide in the alias peaks.

The power spectrum of $\theta^2$ Tau was computed from frequency
values of zero to the Nyquist frequency. The highest power levels were
found in the 10 to 16~cd$^{-1}$ region. An additional band of power
was found in the 26 to 30~cd$^{-1}$ region and will be examined in a later section.
The power spectra in the 8 to 16~cd$^{-1}$ region are presented in Fig.~3.
The second panel from the top shows the dominant three frequencies, which have been found in
all other photometric campaigns as well. The next panel shows the power spectrum after
prewhitening of the main three frequencies. Three additional peaks are detected without
any problems.

The power spectrum of the data after prewhitening six frequencies (Data - 6f) presents
a small problem: the dominant peak may be double. The peak at 11.770~cd$^{-1}$, called f$_7$,
is strong and its frequency easily determined. Removal of this mode leaves another peak
at 11.730~cd$^{-1}$ at a lower amplitude. The frequency separation of 0.04~cd$^{-1}$ is
at the limit of frequency resolution of the present data set. This prevents us from
applying the powerful phasing--amplitude test to distinguish between two independent
close frequencies and the artifacts caused by a single frequency with variable amplitudes
(see Breger \& Bischof 2002). Fortunately, it is not necessary here to prove the
existence of a frequency doublet: both frequencies have been seen before in the
WIRE data (Paper III), while the weaker component in the 1994 data was
already detected in the 1986 data (Paper II). We therefore accept the
reality of the two close frequencies, f$_7$ and f$_8$.

Three additional frequencies are found in the 1994 data. Further peaks are below
the statistical threshold (bottom panel of Fig.~3).

The adopted frequencies and amplitudes are shown in Table~2.
The uncertainties in amplitude
were calculated from the standard formulae given in Breger et al. (1999), based on the
assumption of random observational errors and no aliasing. The uncertainties of the
amplitudes of f$_7$ and f$_8$ are larger because these two modes are very
close in frequency.
Because the separation of two frequencies is at the resolution
limit of the present data set, the computed size of the amplitude
of one mode is somewhat affected by the presence of the other mode (the
so-called pumping effect). These amplitudes are therefore marked with a colon.

\begin{table}
\caption{The photometric frequency spectrum of $\theta^2$ Tau during 1994}
\begin{tabular}{lcccc}
\noalign{\smallskip}
\hline
\multicolumn{3}{c}{Frequency} &  \multicolumn{2}{c}{1994 Amplitudes}\\
& & &  $v$ filter & $y$ filter\\
& cd$^{-1}$ & $\mu$Hz &   mmag & mmag\\
\hline
\noalign{\smallskip}
& & & $\pm$0.1 & $\pm$0.1 \\
\noalign{\smallskip}
f1 &	13.230 & 153.1 & 4.8 &	3.7 \\
f2 & 	13.694 & 158.5 & 5.2 &	4.1\\
f3 & 	13.484 & 156.1 & 3.2 &	2.3\\
f4 &	12.397 & 143.5 & 3.7 & 3.1\\
f5 &	14.316 & 165.7 & 3.7 & 2.8\\
f6 &	10.865 & 125.8 & 2.3 & 1.5\\
f7 &	11.770 & 136.2 & 2.3: &	2.4:\\
f8 &	11.730 & 135.8 & 1.2: & 0.9:\\
f9 &	12.177 & 140.9 & 1.7 & 1.0\\
f10 &	12.832 & 148.5 & 1.4 & 1.4\\
f11 &	14.615 & 169.2 & 1.9 &	1.2\\
\noalign{\smallskip}
\hline
\end{tabular}
\end{table}

The new solution fits the observed data well, as shown in Figs.~1 and 2. The good
fit is also demonstrated by the value of the average residual per single measurement
of $\pm$2.7 mmag in $v$ and $\pm$3.2 mmag in $y$.

\section{Comparison with previous studies}

There now exist three large photometric studies of $\theta^2$~Tau: the present study based on
1994 multisite data, the re-analysis of the 1982--1986 multisite campaigns (with emphasis on
the 1986 data, Paper II), and the WIRE satellite study from space (Paper III).
Each photometric study detected between 10 and 12 frequencies.
The excellent agreement between the frequencies found in these independent studies
is remarkable.

\subsection{The main modes: f$_1$ to f$_5$, f$_8$ to f$_{11}$}

These 9 frequencies were detected in all three studies, but with variable
amplitudes. While during the years 1982, 1983 and 1986, the mode at 13.23~cd$^{-1}$
is dominant in each of these years, during 2002 the 13.69~cd$^{-1}$ frequency became
dominant.

\subsection{The mode at 10.86~cd$^{-1}$: f$_6$}

The mode at 10.86~cd$^{-1}$ is present with amplitudes in excess of 1.0 mmag
in both the 1994 and 2000 data, but appears to be absent ($y$ amplitude $\leq$ 0.3 mmag)
during 1986.

\subsection{The 11.73/11.77~cd$^{-1}$ doublet: f$_7$ and f$_8$}

The frequency separation of this close frequency pair is near the frequency resolution
given by the 25d length of the 1994 data set as well as that of the 1986 and the
WIRE results. In the WIRE data, the two frequencies are both present with similar amplitudes. During
1994, 11.77~cd$^{-1}$ is dominant but the other peak is also visible. In the 1986 data,
we can only see the other, lower-frequency mode. The 1982/3 data offer higher frequency
resolution, but are not useful for the study of the pair because of a higher noise level.

We conclude that both modes are probably present. The strong amplitude variability of these
modes may be intrinsic to the star or an artefact caused by poor frequency resolution.

\subsection{Additional frequencies at 12.70 and 13.81~cd$^{-1}$}

Two additional frequencies are seen at statistically significant levels in only one of the
three data sets:

(i) The 1986 data reveal an additional mode at 13.81~cd$^{-1}$,
which has not been seen before. The power spectrum of the WIRE
residuals shows a peak at 13.83 cd$^{-1}$ (see Fig.~6 of Paper III)
and is therefore probably also present in that data set.

(ii) A mode at 12.70~cd$^{-1}$ was also found by WIRE, which is not affected by 1~cd$^{-1}$
aliasing and is therefore not affected by another mode near 11.73~cd$^{-1}$.
A formal solution with the 1994 data including this frequency yields
a miniscule amplitude of 0.3 mmag. The mode may be present in the 1986 data,
because in the power spectrum a small peak at an amplitude level of $\sim$ 0.5 mmag is seen at
that frequency value, but below the level of significance. Numerical simulations suggest
that the slight 1~cd$^{-1}$ aliasing still present in the multisite ground-based campaign data
is not responsible for the computed small (or zero) amplitudes in 1986 and 1994.

\subsection{Amplitude ratios and phase differences}

The variability measurements from Lowell and Sierra Nevada Observatories were obtained with
the two $v$ and $y$ filters of the $uvby\beta$ system in order to compare
amplitude ratios and phase differences in different colors. Especially the
phase differences for different pulsation modes allow a discrimination between
different $\ell$ values, but extensive measurements of extremely
high photometric accuracy are required.
The present study, regrettably, is not large enough for reliable mode
identifications. In fact, the formal uncertainties in the phase differences
are less than $\pm$3.0$\degr$ for only three modes: f$_1$, f$_2$ and f$_4$.
All three modes give similar amplitude ratios with an average value of
A$_v$/A$_y$ = 1.25 $\pm$ 0.05. For these modes the phase differences,
$\phi_v-\phi_y$, are -7$\degr$ with uncertainties of $\pm2\degr$ for the first
two modes and $\pm3\degr$ for f$_4$.

\section{The high-frequency region}

An additional frequency region with power in excess of the expected noise
occurs in the 26 to 30~cd$^{-1}$ region. This is shown for the 1994
data in the top panel of Fig.~4. Several promising peaks exist, but
do not reach the amplitudes required for a statistically certain discovery of a new
pulsation frequency, viz., amplitude signal/noise ratios $\geq$ 4.0.

The statistics can be improved considerably by examining other
photometric data in the literature, viz. the data obtained during
1982, 1983 and 1986 (see Paper I). For the 1982/1983 and
1986 photometry, we have prewhitened the variability in the low-frequency
domain and computed power spectra in the high-frequency region. This is
also shown in Fig.~4. The peak at 26.2~cd$^{-1}$ is present in all data sets. In spite
of the small amplitude of 0.6 mmag, the detections are statistically significant
in two of the three data sets.
An additional peak at 26.7~cd$^{-1}$ is also statistically significant in the
1982--1986 data, but not seen in 1994. On the other hand, the most promising peak in 1994,
28.9~cd$^{-1}$, is not seen in the earlier data. Since its detection is
not statistically significant in any (including a combined 1982 -- 1994) data
set, it is not adopted by us.

It is interesting that the 26.2~cd$^{-1}$ variability was also seen independently
by the WIRE satellite: Poretti et al. in Paper III ascribe this to a spurious term
caused by the duty cycle and experimental errors. In view of the present detection of the same
frequency in earth-based data, we can rule out an experimental origin and conclude
that these high frequencies originate in $\theta^2$~Tau.

Table~3 lists the amplitudes for the two significant modes.

\begin{figure}
\includegraphics*[width=85mm,clip]{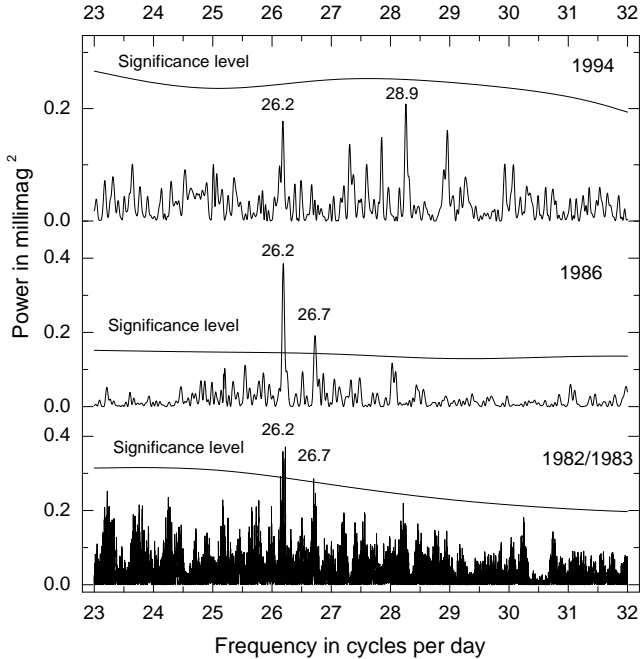}
\caption{Power spectra in the high-frequency region. The
frequency at 26.2~cd$^{-1}$ is present in all data sets, while 26.7~cd$^{-1}$ is
present in both the 1986 and 1982/3 data sets.}
\end{figure}

\begin{table}
\caption{Variability of $\theta^2$ Tau at high frequencies}
\begin{tabular}{lllcccc}
\noalign{\smallskip}
\hline
\multicolumn{3}{c}{Frequency} & \multicolumn{4}{c}{Amplitudes$^1$}\\
& & &  1982/83 & 1986 & 1994 $v$ & 1994 $y$\\
& cd$^{-1}$&$\mu$Hz &  mmag & mmag & mmag & mmag\\
\noalign{\smallskip}
\hline
& & & $\pm$0.1 & $\pm$0.1 &$\pm$0.1 & $\pm$0.1 \\
f12	& 26.184 & 303.1 & 0.6 & 0.6 & 0.5 & 0.4 \\
f13 & 26.732 & 309.4 & 0.5 & 0.5 & 0.1 & 0.1 \\
\noalign{\smallskip}
\hline
\end{tabular}
$^1$~Note that an amplitude of 0.1 mmag does not represent a
statistically significant detection.
\end{table}

\subsection{Could the high frequencies be combination or harmonic frequencies?}

$\delta$~Scuti stars often show combination frequencies, f$_i\pm$f$_j$, as well as slightly
nonsinusoidal light curves leading to 2f$_i$ peaks in the power spectrum. These peaks are,
of course, situated at high frequencies and are
characterized by amplitudes smaller than a factor of ten (or more) relative to those of f$_i$.
Since $\theta^2$~Tau has its main variability in the 11--15~cd$^{-1}$ range, combination frequencies/2f
terms would be found in the 22--30~cd$^{-1}$ region and therefore provide a potential explanation.

The frequencies of the potential high-frequency candidates can be easily calculated from
f$_i\pm$f$_j$ and 2f$_i$. Such have been detected even in low-amplitude $\delta$~Scuti stars
such as 4~CVn (Breger et al. 1999) and XX Pyx (Handler et al. 1996, 2000).

The value of the main high frequency, f$_{12}$ at 26.184~cd$^{-1}$, cannot be identified with the
frequency of a combination of any two modes listed in Table 1. We conclude that it is not
a combination frequency. On the other hand, f$_{13}$ at
26.732 cd$^{-1}$, is in the vicinity of two potential combinations,
f$_1$+f$_3$ = 26.714 and f$_4$+f$_5$ = 26.713~cd$^{-1}$.
Although the 1982--1986 data have an excellent frequency resolution due to
the 1000+ day span, the data contain long gaps, leading to the possibility
of spectral leakage. We can see smaller peaks near 26.714~cd$^{-1}$, so that
some of the power at 26.732 cd$^{-1}$ may
be spectral leakage from combination frequencies. The important
question concerns the physical origin of the combination
terms, e.g., are these combination frequencies just combination terms or are
new modes excited by resonance at frequencies near those of the
combination frequencies?  Garrido \& Rodriguez (1996), in their study of combination
frequencies, point out that the phasing of these combination terms might allow
us to distinguish between the two possibilities. However, for low-amplitude $\delta$~Scuti
stars, accurate phasing of combination terms are observationally difficult to determine
and the question cannot be examined further at this stage.

Observations of $\delta$~Scuti
stars (and other pulsating variables) with known combination frequencies
show that the amplitudes of the combination
terms are a function of the amplitudes of the two frequencies involved in the combination
terms. We can estimate the size of the combination terms, f$_i$+f$_j$ and 2f$_i$
from the Garrido \& Rodriguez analysis as well as the observations of XX~Pyx, 4~CVn and BI~CMi
(Breger et al. 2002) and calculate expected amplitudes small than 0.1 mmag. This is
an order of magnitude smaller than observed. The amplitude argument, therefore, does not
support an identification of the two frequencies as combination modes.

We conclude that the main peak, f$_{12}$ cannot be identified as a combination
frequency, while such an explanation is unlikely for f$_{13}$.

\subsection{Pulsation in a $\delta$~Scuti binary companion}

$\theta^2$ Tau is a 140.728d binary system with two components of similar temperature.
Peterson, Stefanik \& Latham (1993)
have shown that the secondary is fainter by 1.10 mag and still on the main sequence. Both stars are
situated inside the instability strip. It was shown in Paper I that the dominant
pulsation modes originate in the evolved primary component because the
predicted orbital light-time effects for the primary match the observed
shifts in the light curves of up to several minutes. This confirms an expectation
from the values of the frequencies (10 -- 15~cd$^{-1}$)
which are incompatible with those of main-sequence $\delta$~Scuti stars.

What would be the expected frequencies of pulsation of the main-sequence secondary?
Although $\delta$~Scuti stars pulsate with many simultaneously excited modes of
pulsation, the average time-scale obeys a period--luminosity relation of the form,
M$_v~\sim$~-3.05~$\log$~P (see Breger 2000). 
We can now apply the known magnitude difference between the components to provide a rough
estimate of the frequencies expected for the secondary. The magnitude
difference of 1.10 mag shifts the 11--15~cd$^{-1}$
range of the primary to 25--33~cd$^{-1}$, which agrees with the band of power
seen in the power spectrum.

The light of the secondary is heavily diluted by the light from the primary, which is
1.1 mag brighter than the secondary. Consequently, the amplitudes of pulsation
of the secondary are reduced to one fourth of their intrinsic values.
If our interpretation is correct, then the secondary would
have $V$ amplitudes of about 0.5 mmag in combined light, or 2 mmag
in the star alone. The hypothesis that these modes come from the secondary
can be tested: the orbital motions produce light-time shifts of several minutes.
Regrettably, the observed amplitudes are too small to obtain the required accuracy in the
phasing at different orbital times.

\section{Pulsation models for $\theta^2$~Tau}

In order to examine the nature of the pulsations of $\theta^2$~Tau in
more detail, the physical parameters for the two stars forming the
binary system need to be determined.  A recent summary of different
determinations of global parameters of $\theta^2$~Tau can be found in
de Bruijne et al. (2001).  Torres et al. (1997) used the orbital
parallax to derive absolute visual magnitudes, $M_{\rm{v}}$, of 0.37 and
1.47 for the two components.  Similar values, 0.48 and 1.58, are
obtained by de Bruijne et al. (2001) and Lebreton at al. (2001) from
Hipparcos dynamical parallax.  Both stars have similar
temperatures and its value can be obtained from the Moon \& Dworetsky
(1985) calibration of $uvby\beta$ photometry, viz., $T_{\rm{eff}}$ = 7950\,K.
This value is lower than the value of 8200\,K adopted by Breger et al.
(1987) as well as others, since the older model-atmosphere calibrations
of $uvby\beta$ tended to overestimate the temperatures of A stars.
Using a calibration of the $UBV$ photometry, de Bruijne et al. (2001)
derive $T_{\rm{eff}}$ = 7980\,K for the primary 
and $T_{\rm{eff}}$ = 8230\,K for the secondary component

The Moon \& Dworetsky (1985) calibration also allows us to derive a
$\log~g$ = 3.69 from the $uvby\beta$ photometry of the combined light
of the two stars. Correction of the measured Balmer discontinuity,
$c_{\rm{1}}$, for the less evolved secondary places our estimate for
the primary near $\log~g$ = 3.55.

This implies a considerably higher evolutionary status than that
deduced by Torres et al., who adopted $\log~g$ = 4.0. However, our
value is consistent with the position of the star in the
Hertzsprung--Russell Diagram for the Hyades cluster as well as the
known luminosity (see above).  We can also derive the $\log~g$ value
from the known mass, luminosity and temperature: for a mass of 2.42
solar masses for the primary (Torres et al.) a $\log~g$ value of 3.63
is derived, in agreement with the value found above from $uvby\beta$
photometry.

Finally, this as well as similar studies neglect the effect of rotation
and aspect, which affect the choice of the appropriate values for
temperature, luminosity and gravity (e.g., see P\'erez Hern\'andez et al. 1999).

\subsection{Method of computation}

The method of computation of stellar evolution and pulsation was the
same as in our other studies of $\delta$ Scuti stars (see Pamyatnykh
2000 and references therein). In particular, we used the latest version
of the OPAL opacities (Iglesias \& Rogers 1996) supplemented with the
low--temperature data of Alexander \& Ferguson (1994).  Also, we used
an updated version of the OPAL equation of state (Rogers et al. 1996),
viz., version EOS2001, which was copied from the OPAL data
library\footnote 
{ 
{\tt http://www-phys.llnl.gov/Research/OPAL/opal.html
ftp://www-phys.llnl.gov/pub/opal/eos2001/} 
}.

The computations were performed starting with chemically uniform models
on the ZAMS, assuming an initial hydrogen abundance $X=0.716$, a helium
abundance $Y=0.26$ and a heavy element abundance $Z=0.024$. These values
correspond to recent estimates of the chemical composition of the
Hyades (Perryman et al. 1998, Lebreton et al. 2001). 
We also tested models of somewhat different chemical composition,
with slightly higher metallicity or with higher helium content
($X=0.723$, $Y=0.25$, $Z=0.027$; $X=0.686$, $Y=0.29$, $Z=0.024$).  

We computed models with and without overshooting
from the convective core. In the first case, the overshooting distance,
$d_{\rm{over}}$, was chosen to be $0.2 \, H_{\rm{p}}$, where
$H_{\rm{p}}$ is the local pressure scale height at the edge of the
convective core.  Examples of evolutionary tracks for $\delta$~Scuti
models computed with and without overshooting are given by Breger \&
Pamyatnykh (1998) and Pamyatnykh (2000).
In the stellar envelope the standard mixing-length theory of convection
with the mixing-length parameter $\alpha$ = 1.6 was used. This value
was also used by Perryman et al. (1998) and Lebreton et al. (2001), who derived
$\alpha$ from the calibration of the solar-model radius. 

Uniform (solid-body) stellar rotation and conservation
of global angular momentum during evolution
from the ZAMS were assumed for all computations. These assumptions were chosen
due to their simplicity. The influence of rotation on the
evolutionary tracks of $\delta$~Scuti
models was demonstrated by Breger \& Pamyatnykh (1998) and
Pamyatnykh (2000). In most cases the initial equatorial
rotational velocity on the ZAMS was chosen to be 100 km/s.

The linear nonadiabatic analysis of low-degree oscillations
($\ell$\,$\leq$\,2) was performed using the code developed by
Dziembowski (1977).  The effects of slow rotation on oscillation
frequencies were treated up to third order in the rotational velocity
(Dziembowski \& Goode 1992, Soufi et~al. 1998).

The main aim of the present theoretical study
was to check the instability of models of both stellar components of $\theta^2$ Tau
for the whole range of allowed values of mass or gravity, effective
temperature and luminosity and to outline possible asteroseismological
constraints to the models. We did not attempt to construct any model which
fits quantitatively the observed and the theoretical frequency spectra
of excited multiperiodic oscillations. 

As a first step, we tried to fit the observed and theoretical
frequency ranges of the unstable modes. The main uncertainty of such a
study is the unsatisfactory description of
convection in the stellar envelope and its interaction with pulsations.
The fact that convection may influence the instability of stars not
only near the Red Edge but almost in the whole classical instability strip
was discussed by Stellingwerf (1984) in connection with the problem
of the determination of helium abundance in globular clusters from the
temperature at the Blue Edge of the RR Lyrae instability domain.
$\delta$ Scuti stars are generally hotter than RR Lyrae stars, but even
in this part of the instability strip convection influences the
onset of the instability for low overtones of radial and
nonradial oscillations.  Fig.~9 of Pamyatnykh (2000) demonstrates
that convection significantly displaces
the radial fundamental Blue Edge to hotter temperatures due to
increased driving in the hydrogen ionization zone which is more
extended than in the radiative models.
A part of this contribution to
the driving may also be caused by the generally-used assumption that the
convective flux does not vary during an oscillation cycle. This
simple assumption is probably incorrect in the hydrogen convection zone.
Therefore, the reality of the additional convective destabilization in
the outer hydrogen zone must be examined by using a nonlocal
time-dependent treatment of convection. First promising
results in this direction (see Michel et al. 1999, Houdek 2000) exist. Note
that the total driving in a $\delta$~Scuti star (with main contribution
due to the $\kappa$ mechanism operating in the second helium ionization zone)
only slightly exceeds the total damping. Therefore even small
contributions to the driving are important. 

Due to these uncertainties we consider our results as preliminary and
our models as test models. Nevertheless, even with
such a simple description of the convection we may have a possibility
to constrain the stellar parameters from instability studies.

\subsection{Results for test models}

\begin{figure}
\centering
\includegraphics*[width=88mm]{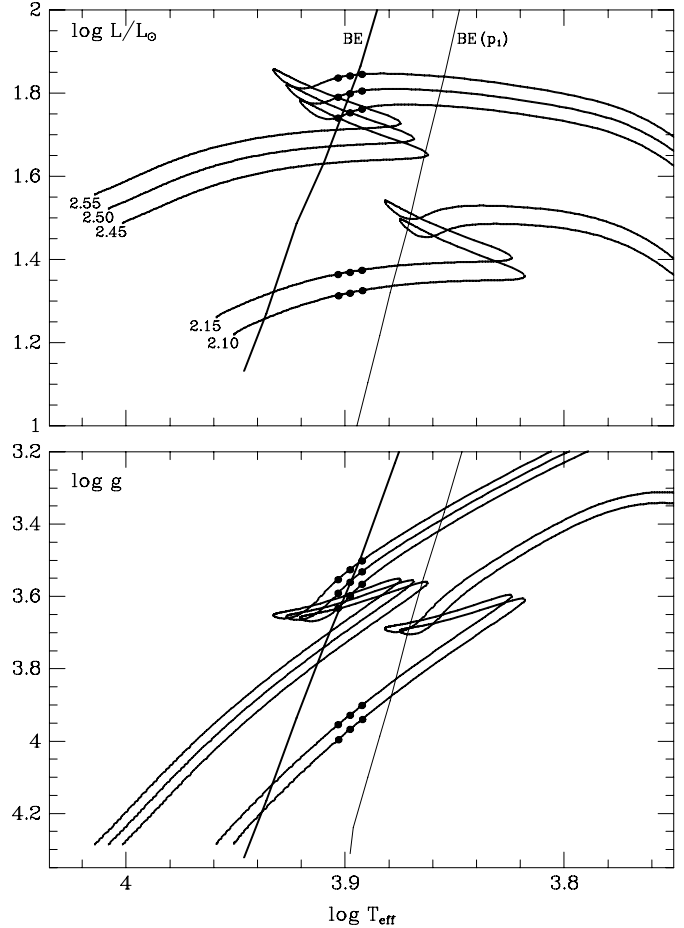}
\caption{
Evolutionary tracks for indicated values of $M / M_{\odot}$.  Test
models with effective temperature $T_{\rm{eff}}=8000$, 7900 and 7800\,K
for the primary and the secondary component of $\theta^2$~Tau
are marked by filled circles. All models were computed for an initial
hydrogen abundance $X=0.716$, a helium abundance $Y=0.26$ and a heavy
element abundance $Z=0.024$. A mixing-length parameter of convection
$\alpha=1.6$ was used. Overshooting from the stellar convective
core was taken into account with an overshooting distance
$d_{\rm{over}} = 0.2 \, H_{\rm{p}}$. The equatorial rotational
velocity on the ZAMS was chosen to be 100~km/s.  The almost vertical
lines show the general Blue Edge of the instability
strip and the Blue Edge of the radial fundamental mode, $p_1$,
as computed for a standard chemical composition (Pamyatnykh 2000).
}
\end{figure}

In Fig.~5 we show evolutionary tracks for selected models, computed by
taking overshooting from the stellar convective core into account.  The
models with effective temperature $T_{\rm{eff}}=8000$, 7900 and 7800\,K,
in which we study the oscillations, are marked by filled circles.  The
luminosities of the higher and the lower-mass models approximately 
correspond to the luminosity of the primary and secondary stellar components,
respectively.

For the primary component, we have chosen models at the
beginning of the post-main-sequence (hereafter called post-MS)
expansion stage, which burn
hydrogen in a thick shell just after full hydrogen exhaustion
in the stellar core. This evolutionary stage of the primary was
proposed earlier by Kr\'olikowska (1992) for a significantly hotter
model of about 8200 K without overshooting.  Moreover, we studied
slightly more massive models of similar effective temperature and
luminosity in the MS stage as well as post-MS models without
overshooting.

The secondary component, which is fainter by approximately 1.1 mag (which
corresponds to the difference by 0.44 in $\log L$ between the components)
is a main-sequence star with hydrogen burning in the convective core.

In Fig.~5 we also plot the Blue Edges of the instability strip taken from
Pamyatnykh (2000), which were computed for nonrotating models without
overshooting for $X=0.70$, $Y=0.28$, $Z=0.02$ and with an assumed mixing-length
parameter $\alpha = 1.0$. The general Blue Edge is defined as the hotter
envelope of unstable overtones, from radial mode $p_8$ (seventh overtone) near
the ZAMS to mode $p_5$ (fourth overtone) for highest luminosities or
smallest gravities in the figure (see Fig.~3 in Pamyatnykh 2000).
The Blue Edge for the radial fundamental mode lies approximately in the
center of the $\delta$~Scuti instability strip. 

An additional study of the instability along the 2.5 $M_{\odot}$
evolutionary track shows that the best theoretical general Blue Edge for $X=0.716$,
$Y=0.26$, $Z=0.024$ will be located very close to the blue edge shown in Fig.~5, 
because the differences in the rotational velocity and in the
overshooting efficiency do not influence the position of the Blue Edges
and because the differences in the helium abundance are small.
Moreover, convection has only a minor influence on the position of this
hot general Blue Edge (see Fig.~9 in Pamyatnykh 2000).
For the fundamental radial mode the best Blue Edge will be hotter by
$0.008-0.009$ in $\log T_{\rm{eff}}$. This is mainly due to a higher value of
the mixing-length parameter.

From Fig.~5 we immediately obtain a strong constraint on the possible
effective temperature of the primary of $\theta^2$~Tau. All models 
with $\log (L/L_{\odot})>1.67$ and $\log T_{\rm{eff}}>3.907$
($T_{\rm{eff}}>8070$\,K) are stable in all modes. 
A MS model of 2.5 $M_{\odot}$ on the Blue Edge 
($T_{\rm{eff}} = 8070$\,K) is marginally unstable in radial mode $p_6$ with
the frequency 18.74~cd$^{-1}$, which is well outside the observed
frequency range.  We can conclude that only significantly cooler models
can pulsate with the observed frequencies in the 10.8 to 14.6~cd$^{-1}$
range.

\begin{figure}
\centering
\includegraphics*[width=88mm]{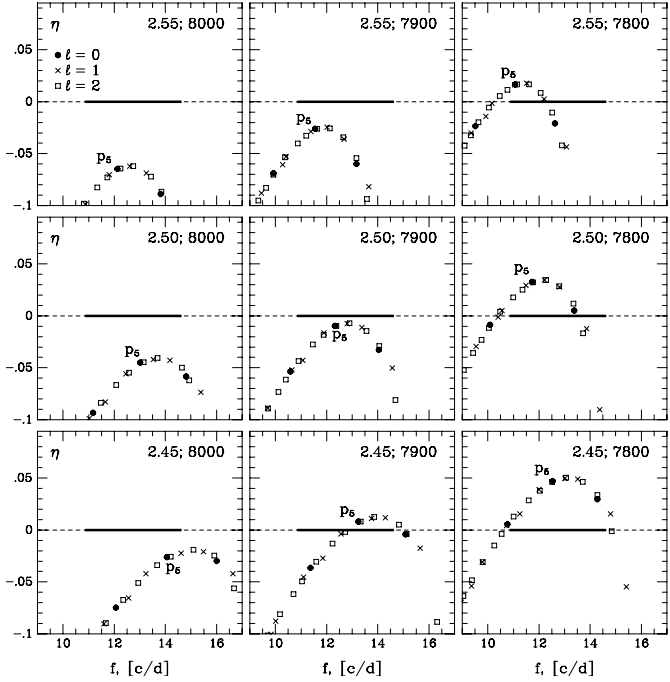}
\caption{
  Normalized growth rates, $\eta$, plotted against frequency, $f$, in
test models of the primary component of $\theta^2$~Tau. Positive values
of $\eta$ correspond to unstable modes. The values of
$M$ (in solar units) and $T_{\rm eff}$ are given in each panel.  The
position of the models in the HR diagram is shown in Fig.~5.  Models
were calculated with initial abundances $X=0.716$, $Y=0.26$,
$Z=0.024$ and $V_{\rm{rot}}$(ZAMS) = 100 km/s.  The symbol $p_5$ is
plotted near the corresponding radial overtone.  The thick horizontal
line shows the range of frequencies (10.865 to 14.615~cd$^{-1}$) observed for
the primary component of $\theta^2$~Tau.
}
\end{figure}

This conclusion is confirmed by computation of oscillations of the
selected test models for the primary of $\theta^2$~Tau. In Fig.~6, the
normalized growth rates of radial and nonradial modes are plotted
against frequency for all nine higher-mass models which are marked in
Fig.~5.  Only axisymmetric modes ($m = 0$) are shown.  The independence
of the growth rate on the spherical harmonic degree, $\ell$, is a
typical feature of modes excited by the $\kappa$ mechanism.  The
rotational velocities of the models are 81 to 86 km/s. The rotational
splitting of the modes can extend the frequency range by
approximately 0.5 c/d on both sides.  We can see that excited
frequencies of the 2.45 $M_{\odot}$ model with $T_{\rm{eff}} = 7800$\,K
are in excellent agreement with the observed frequency range.  The
frequency range of unstable modes spans three radial orders from $p_4$ to
$p_6$ for radial modes (mode $p_4$ is marginally unstable).

\begin{figure}
\centering
\includegraphics*[width=88mm]{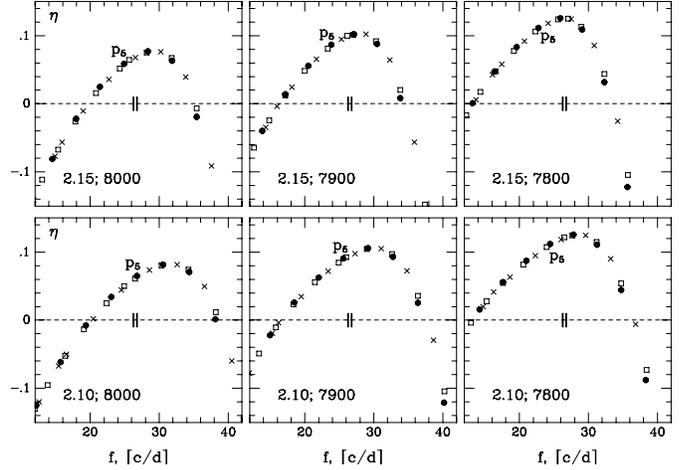}
\caption{
The same as in Fig.~6 but for test models of the secondary component.
Short vertical lines mark the observed frequencies 26.184 and 26.732~cd$^{-1}$.
}
\end{figure}

As was noted already, the results are sensitive to the treatment of convection.
For example, if we use a mixing-length parameter $\alpha=2.0$ instead of
$\alpha=1.6$, the frequency range of unstable modes for 2.45 $M_{\odot}$ model
with $T_{\rm{eff}} = 7800$\,K is extended by 1 c/d on both sides. Moreover,
our assumption about the unperturbed convective flux during an oscillation
cycle is not fulfilled inside the hydrogen convective zone and may result
in artificial additional driving in this zone. Therefore, these preliminary
results must be considered with caution. 

Similar results were obtained for the models without overshooting,
the best fitting is achieved in this case for 
2.50~$M_{\odot}$ model with $T_{\rm{eff}} = 7800$\,K.
Also, for slightly more massive MS models with overshooting,
we obtained a good agreement between the observed and the theoretical
frequency ranges. The parameters of some models are given in Table 4 below.

In Fig.~7 we show the normalized growth rates in test models of the secondary
component. These models are also marked in Fig.~5. As for the primary,
we used models with effective temperatures from 7800 to 8000\,K. 
The frequency range of unstable modes is much wider for these less
luminous models: it spans 5 to 7 radial orders from $p_2$ to $p_8$.
The observed frequencies lie near or between radial modes $p_5$ to $p_6$,
which are unstable in the whole range in effective temperature.

\begin{table*}
\begin{center}
\caption
{
Parameters of the best test models of $\theta^2$~Tau which fit the
observed frequency ranges. Models of 2.65, 2.50 and 2.45 $M_{\odot}$
correspond to the primary component, models of 2.10 $M_{\odot}$
correspond to the secondary component.
The symbols have their usual meanings (see text). The last column gives
the frequency range, in cycles per day, of unstable modes for each model.
}
\begin{tabular}{|cccccccccc|}
\hline
Model & $M/M_{\odot}$ & Ev. Stage & $Age\,[Myr]$ & $T_{\rm eff}$ & $\log L$
& $\log \, g$ & $V_{\rm rot}$ & $d_{\rm over}$
& Frequency range\\
\hline
P1 & 2.65 &    MS   &  535 & 7800 & 1.792 & 3.58 & 60 & 0.2 & 10.7 -- 15.4\\
P2 & 2.50 & post--MS & 595 & 7800 & 1.780 & 3.56 & 83 & 0.0 & 10.5 -- 14.6\\
P3 & 2.45 & post--MS & 695 & 7800 & 1.762 & 3.57 & 84 & 0.2 & 10.6 -- 14.9\\
\hline
S1 & 2.10 &   MS     & 667 & 7900 & 1.319 & 3.97 & 91 & 0.2 & 13.0 -- 37\\
S2 & 2.10 &   MS     & 704 & 7800 & 1.325 & 3.94 & 90 & 0.2 & 16.5 -- 38\\
\hline
\end{tabular}
\end{center}
\end{table*}

In Table 4 we summarize parameters of the best test models.  We can see
that both MS and post--MS models for the primary with
$T_{\rm{eff}} = 7800$\,K can fit the observed frequency range.
The values of the surface gravity and luminosity of all three models are similar.
In the computation of the evolution of the MS model P1, the initial equatorial
rotational velocity on the ZAMS was chosen to be 80 km/s.
(Rotation does not affect our theoretical results significantly, but it may
influence the photometric calibrations.)
The total number of unstable modes of $\ell \leq 2$ is
equal to 37 in the MS model P1, 77 and 56 in the post-MS models P2 and
P3, respectively. 

Note that the post-MS evolution of models for the primary is much faster
than the MS evolution of the less massive secondary. For example, in
Fig.~5 all models of 2.45 $M_{\odot}$ with $T_{\rm{eff}} = 8000$\,K,
7900\,K and 7800\,K (the last model is just model P3) have the same age
695 Myr. These models correspond to stars which are slightly older than the Hyades
(650 Myr, Lebreton et al. 2001). On the other
hand, the models without overshooting may correspond to stars younger than the Hyades:
the age of 2.50 $M_{\odot}$ models is 595 Myr. The best model for
the primary may therefore be one with slightly weaker overshooting than that of
model P3.

The inferred age of the secondary star can be adjusted to the age of the
primary as well as to that of the Hyades cluster. This is possible
because of the slower main-sequence evolution of the secondary and due to the fact
that corresponding models of different effective temperatures fit the
observed frequencies quite well. As an example, models of 2.10 $M_{\odot}$
with $T_{\rm{eff}} = 8000$\,K, 7900\,K and 7800\,K  have the ages 625, 667 and 704 Myr.

Moreover, we tested some models of different chemical composition.
If we choose the estimated values for the helium content (Lebreton et al. 2001)
and metallicity (Lastennet et al. 1999), the chemical compositions
become $X=0.723$, $Y=0.25$, and $Z=0.027$.
The results for these models are very similar to those computed above for
$X=0.716$, $Y=0.26$, $Z=0.024$.

Models with $X=0.686$, $Y=0.29$, $Z=0.024$ are somewhat more luminous
for a given mass, e.~g., the track of 2.45~$M_{\odot}$ almost
coincides with the track of 2.55 $M_{\odot}$ computed with
$X=0.716$, $Y=0.26$, $Z=0.024$. The models are also pulsationally more unstable and
the range of unstable modes is somewhat wider due to higher helium 
abundance. As a consequence, the MS model of 2.45 $M_{\odot}$ with 
$T_{\rm{eff}} = 7900$\,K can be unstable in the observed frequency
range and also at slightly higher frequencies. The luminosity
of this model, $\log (L/L_{\odot})=1.71$, agrees with the
estimates from the orbital and dynamical parallaxes.

\section{Conclusion}

The DSN campaign~12, carried out during 1994 at four observatories, resulted
in 152 hours of high-precision photometry. 13 frequencies of pulsation were derived
from the data. These frequencies confirm the
results from previous Earth-based (1982--1986) as well as satellite (2000) photometry,
although amplitude variability on a time scale of several years is present.

While the detected modes in the 10--15 cd$^{-1}$ range probably all
originate in the primary, the higher frequencies with small amplitudes
probably originate in the $\delta$~Scuti-type of variability of the main-sequence
companion.

We constructed evolutionary models both for the primary and the
secondary components and tested them for the instability
against radial and nonradial oscillations. 

The best fit of the theoretical and observed frequency ranges is
achieved for models with $T_{\rm{eff}}\approx 7800$\,K or slightly
higher, in agreement with photometric calibrations. The instability
spans two or three radial orders from $p_4$ to $p_6$ for radial modes.
Post-MS models with or without overshooting are preferable for the primary, but
MS models with overshooting are also possible.

For the less luminous secondary component the instability range is wider
and spans 5 to 7 radial orders from $p_2$ to $p_8$. Models of different
effective temperature are unstable at observed frequencies 
26.184 and 26.732~cd$^{-1}$ which lie around radial modes $p_5$ to $p_6$.

The main uncertainties of the results are caused by our crude treatment of
the convective flux.  We neglected the Lagrangian variation of the
convective flux during an oscillation cycle. This assumption is not
fulfilled in the hydrogen convective zone, which may lead to an incorrect
estimate of the small contribution of this zone to the driving.
To prove the results quantitatively, it is necessary to use a reliable
theory of non-local time-dependent convection. We note also the nonlocal
results by Kupka \& Montgomery (2002), who point out the necessity
to use very different values of the mixing-length parameter in the
hydrogen and helium convection zones, if we still use local
mixing-length theory. According to nonlocal studies of A-star
envelopes, it is necessary to use a small value of this parameter in the
hydrogen zone and a significantly higher value in the deeper helium
zone. We plan to perform corresponding tests for the $\theta^2$ Tau
models. However, even with the present calculations and results
we probably can exclude modes involving
the second radial overtone (mode $p_3$) and lower-order modes, as well as
all hot models of the primary with $T_{\rm{eff}} > 8000$\,K. The
reason is that all these models are stable in the observed frequency range
of 10.8 to 14.6~cd$^{-1}$.

\section*{Acknowledgements}
It is a pleasure to thank F. Beichbuchner for assistance with the observations.
We are grateful to W. A.~Dziembowski for stimulating discussions. This investigation
has been supported by the Austrian Fonds zur F\"{o}rderung der wissenschaftlichen
Forschung under project number P14546-PHY (MB) and the Polish Committee for
Scientific Research under Grant 5-P03D-012-20(AAP).


\begin{thebibliography}{}

\bibitem{} Alexander~D.~R., Ferguson~J.~W., 1994, ApJ, 437, 879
\bibitem{} Breger, M., 1993, in Butler~C.~J., Elliott~I., eds,
Stellar Photometry - Current Techniques and Future Developments,
Cambridge University Press, 106
\bibitem{} Breger M., 2000, in Breger M., Montgomery M.~H., eds,
Delta Scuti and Related Stars, ASP Conf. Ser. 210, 3
\bibitem{} Breger M., 2002, Communications in Asteroseismology (Vienna), 141, 3 (Paper II)
\bibitem{} Breger M., Bischof K.~M., 2002, A\&A, in press
\bibitem{} Breger~M., Pamyatnykh~A.~A., 1998, A\&A, 332, 958
\bibitem{} Breger M., Huang L., Jiang S.-y., Guo Z.-h., Antonello E.,
Mantegazza L., 1987, A\&A, 175, 117
\bibitem{} Breger M., Garrido R., Huang L., Jiang S.-y., Guo Z.-h., Frueh M.,
Paparo M., 1989, A\&A, 214, 209 (Paper I)
\bibitem{} Breger M., Stich J., Garrido R., et al., 1993, A\&A 271, 482
\bibitem{} Breger M., Handler G., Garrido R., et al., 1999, A\&A, 349, 225
\bibitem{} Breger M., Garrido R., Handler G., et al., 2002, MNRAS, 329, 531
\bibitem{} de Bruijne J. H. J., Hoogerwerf R., de Zeeuw P. T., 2001,
  A\&A, 367, 111
\bibitem{} Duerbeck H. W., 1978, IBVS, 1412, 1
\bibitem{} Dziembowski~W.~A., 1977, Acta Astron., 27, 95
\bibitem{} Dziembowski~W.~A., Goode~P.~R., 1992, ApJ, 394, 670
\bibitem{} Ebbighausen E. G., 1959, Pub. Dom. Astrophys. Obs., 11, 235
\bibitem{} Garrido R., Rodriguez E., 1996, MNRAS, 281, 696
\bibitem{} Handler, G., et al., 1996, A\&A, 307, 529
\bibitem{} Handler, G., et al., 2000, MNRAS, 318, 511
\bibitem{} Horan S., 1977, IBVS, 1232, 1
\bibitem{} Horan S., 1979, AJ, 84, 1770
\bibitem{} Houdek~G.~A., 2000, in Breger M., Montgomery M.~H., eds,
Delta Scuti and Related Stars, ASP Conf. Ser. 210, 454
\bibitem{} Iglesias~C.~A., Rogers~F.~J., 1996, ApJ, 464, 943
\bibitem{} Kennelly E. J., Walker G. A. H., 1996, PASP, 108, 327
\bibitem{} Kovacs G., Paparo M., 1989, MNRAS, 237, 201
\bibitem{} Kr\'olikowska M., 1992, A\&A, 260, 183
\bibitem{} Kupka F., Montgomery M.~H., 2002, MNRAS, 330, L6
\bibitem{} Lastennet E., Valls-Gabaut D., Lejeune Th., Oblak E.,
  1999, A\&A, 349, 485
\bibitem{} Lebreton Y., Fernandes J., Lejeune T., 2001, A\&A, 374, 540
\bibitem{} Li Z.-p., Zhou A.-y., Yang D., 1997a, Acta Astrophys. Sinica, 17, 166
\bibitem{} Li Z.-p., Zhou A.-y., Yang D., 1997b, PASP, 109, 217
\bibitem{} Michel E., Hern\'andez M. M., Houdek G., et al.,
  1999, A\&A, 342, 153
\bibitem{} Moon T.~T., Dworetsky M. M., 1985, MNRAS, 217, 305
\bibitem{} Pamyatnykh~A.~A., 2000, in Breger M., Montgomery M.~H., eds,
Delta Scuti and Related Stars, ASP Conf. Ser. 210, 215
\bibitem{} P\'erez Hern\'andez F., Claret A., Hern\'andez~M.~M., Michel~E., 1999, A\&A, 346, 586
\bibitem{} Perryman M. A. C., et al., 1998, A\&A, 331, 81
\bibitem{} Peterson D. M., Stefanik R. P., Latham D. W., 1993, AJ, 105, 2260
\bibitem{} Poretti E., Buzasi D., Laher R., Catanzarite J., Conrow T.,
2002, A\&A, 382, 157 (Paper III)
\bibitem{} Rogers~F.~J., Swenson~F.~J., Iglesias~C.~A., 1996, ApJ, 456, 902
\bibitem{} Soufi~F., Goupil~M.-J., Dziembowski~W.~A. 1998, A\&A, 334, 911
\bibitem{} Sperl M., 1998, Communications in Asteroseismology (Vienna), 111, 1
\bibitem{} Stellingwerf R.~F., 1984, ApJ, 277, 322
\bibitem{} Torres G., Stefanik R. P., Latham D. W., 1997, ApJ, 485, 167
\end{thebibliography}
\end{document}